\documentclass[12pt,a4paper,amsmath,amssymb,apsmy]{revtex4}

\usepackage{amsfonts}
\usepackage{graphics} 

\begin{document}

\title{Spherical Gravitational Waves in Relativistic
Theory of Gravitation }

\author{A.A. Leonovich} \affiliation{
 University of informatics and Radioelectronics, Minsk, Belarus},
\author{Yu.P. Vyblyi}
\affiliation{
 Institute of Physics, National Academy
of Sciences of Belarus, Minsk,Belarus
\\e-mail:Vyblyi@gmail.com}

\begin{abstract}{Within the framework of relativistic theory of gravitation the exact
spherically-symmetric wave solution is received. It is shown that
this solution possesses the positive-definite  energy and momentum
deriving with the  Fock energy-momentum density tensor of
gravitational field.  In this connection the sense of Birkhoff theorem in
Relativistic Theory of Gravitation  is discussed.}
\end{abstract}

\maketitle

\section
{SPHERICALLY-SYMMETRIC WAVE SOLUTION}

In General Relativity the energy of gravitational field is described
with help the energy-momentum pseudotensor, but the expression for
corresponding tensor is absent. It is one of reasons for regarding the
gravitation interaction  as a tensor interaction in
Minkowski space-time (see, for example \cite{Fey}). The most
consistently such approach was realized in the so-called  relativistic
theory of gravitation (RTG) \cite{Log1}, \cite{Log2}. This theory can
be considered  as a gauge theory of the group of Lie variations for
dynamical variables. The related transformations are variations of
the form of the function for generally covariant transformations.
In order to  the action be invariant for this group under the transformations
of the dynamic variables alone one needs  replacing the ''nondynamic''
Minkowski metric $\gamma ^{ik}$
with expression $g^{ik}$: $\tilde{g}^{ik}=\sqrt{-g}g^{ik}=\sqrt{-\gamma }%
(\gamma ^{ik}+k\psi ^{ik})$, where $\gamma =det\gamma _{ik},g=detg_{ik}$, $%
k^{2}$ - is the Einstein constant, and thus introducing the gauge
gravitational potential $\psi ^{ik}$. The quantity  $g^{ik}$ is
interpreted here as a metric of an effective space-time allowing an unique construction
of the  connection (the Cristoffel bracketS). The RTG field equations at its massless variant are
the Einstein ones for this effective metric, added the conditions,
restricting the spin states of the field $\psi ^{ik}$
\begin{equation}
D_i\tilde g^{ik}=0~,
\end{equation}
 where $D_i$ - the covariant derivative in the Minkowski space. This
condition plays the significant role in RTG. It removes the gauge
arbitrariness of Einstein equations and coincides with the Fock
harmonical condition in Galilean coordinates \cite{Fock}.

Although the massless RTG field equations locally coincide with General Relativity ones,
 their  global solutions, generally speaking, will be different,
since this solutions are defined on the various manifolds.  RTG,
being founded on the simple space-time topology, allows to  introduce the
global Galilean coordinate system, that distinguishes RTG from the
bimetric theories, in which a flat space plays the auxiliary role and
its topology does not  define the character of the physical processes.
This  distinction may take place at interpretation of the field
solutions, since the coordinate system in RTG is defined by Minkowski
metric, but it is fixed by noncovariant coordinate conditions in GR.
Just this situations takes place for spherical-symmetric
gravitational fields. In GR according the Birkhoff  theorem
\cite{GRel} any spherical gravitational field in vacuum is  static.
The proof of this theorem is grounded on the transformation of certain
spherically-symmetric metric to the coordinates in which it has a
static form. But in RTG such transformation is the transfer from the
spherical  coordinates in the Minkowski space with the metric
$\gamma_{ik}= diag(1,-1,-r^2, -r^2\sin^2\theta)$ to some "nonstatic"
coordinates.  The Birkhoff theorem means that in the case of
spherical symmetry  the coordinate system in which the vacuum metric
depends from  one coordinate always exists, but it not means that the
field was static in the starting coordinates.

Hence the task of the investigation of nonstatic spherical-symmetric
solutions, which was in general view investigated in \cite{Vlas},
arises. In this paper one of the possible nonstatic
spherically-symmetric wave solution is found in implicit form. Such
solutions may play very important role in astrophysics and cosmology.

To find the spherical wave solutions we use the Birkhoff theorem
and
present a nonstatic spherical vacuum solution in the certain coordinate  system $%
(\tau,R,\theta,\phi)$ in the form of  Schwarzschild metric
\begin{equation}
ds^2=(1-\frac{2m}{R})d{\tau}^2-(1-\frac{2m}{R})^{-1}dR^2-R^2
d\Omega^2~.
\end{equation}

To find the solution in spherical coordinates of
$(t,r,\theta,\phi)$ we make the coordinate transformation
\begin{equation}
t=t(\tau,R), r=r(\tau,R)~,
\end{equation}

\noindent
 The  transformation coefficients will be found from the
condition (1). The corresponding equations  connecting the variables
$(t,r)$ and $(\tau,R)$,  take the form
\begin{equation}
\frac{R}{R-2m}\frac{\partial^2 t}{\partial {\tau}^2}- R^{-2}\partial_R [(R^2- 2mR)%
\frac{\partial t}{\partial R}]=0~,
\end{equation}
\begin{equation}
\frac{R}{R-2m}\frac{\partial^2 r}{\partial {\tau}^2}- R^{-2}\partial_R [(R^2- 2mR)%
\frac{\partial r}{\partial R}]+\frac{2r}{R^2}=0~.
\end{equation}

We  search for the partial solution of the equations (4), (5)
in the next form
\begin{equation}
\tau=t+T(u), R=r+m, u=t+f(r)~,
\end{equation}
where u - the retarded argument, which is finite at any values  of
$r$; the light velocity and gravitational constant are believed equal to
unit.  Another form of solution was considered in \cite{Leo}. Finding
with help of (6) the transformations coefficients and substituting
its to equations (4),(5), we find that the equation (5) is satisfied
identically, and the  equation (4) after variables separation is
reduced to ordinary differential equations with separation constant
$C$ for the functions $T(u)$ and $f(r)$
\begin{equation}
T_{uu} = C T_u (1+T_u)^2~,
\end{equation}
\begin{equation}
f_{rr}+ C{f_r}^2 +\frac{2r}{r^2 - m^2}f_r -
C\frac{(r+m)^2}{(r-m)^2}=0~,
\end{equation}

The first integral of equation (7) has a form
\begin{equation}
Cu = \frac{1}{1+z} - \ln|\frac{z+1}{z}| + A~,
\end{equation}
where $z=T_u$, $A$ - integration constant. From (9) we find that
function $z(u)$ is positive for positive $A$ and large values of
$u$ ($u>0$).

The equation (8) may be integrated with help the power series for
$f(r)$ relative variable $1/r$. The analysis of this solution shows
that for $C>0$, $f_r < 0$ for all values of $r$.
The metric components now have the next form
\begin{equation}
g_{00} = \frac{r-m}{r+m} (1+T_u)^2~,  g_{01} = \frac{r-m}{r+m} T_u (1+T_u)f_r~,%
\end{equation}
\begin{equation}
g_{11} = \frac{r-m}{r+m} {T_u}^2 {f_r}^2 - \frac{r+m}{r-m}~,
g_{22} = -(r+m)^2~,  g_{33} = -(r+m)^2 \sin^2\theta~.
\end{equation}

\section
  {FOCK ENERGY-MOMENTUM TENSOR OF GRAVITATIONAL FIELD}

The finding metric may be used for the investigation of a  sign of
gravitational radiate density $t^{00}$, which according to RTG has a form %
\cite{Log1}

\begin{equation}
t^{00}= \frac{1}{\sqrt {-\gamma}}\gamma^{ik}D_i{D_k}{\tilde
g}^{00}~.
\end{equation}
The problem concerning positive definite of gravitational energy
density is not trivial, because in RTG the expression $t^{00}$ don't
possess square-law structure relative the field functions and its
first derivatives and it contains second derivatives too. This
situation is the consequence of a difference between spin one and
spin two description. In the case of a vector field Lagrangian does
not contain the Christoffel symbols (in arbitrary coordinate system)
and the energy-momentum tensor can be easily found when this
Lagrangian is varied in respect to the metric. The same is true for
the Maxwell and Yang-Mills fields too. Quiet different situation
arises in the case of a tensor field. Now the Christoffel symbols can
not be shorten and thus the energy-momentum tensor contains the
second derivatives. This fact leads,  generally speaking to
uncertainty of energy density sign. The possible approaches to the
solving of this problem are regarded in \cite{Mla},\cite{Leon}.

Let us consider Fock representation for Einstein tensor $G^{ik}$
\cite{Fock}

\begin{equation}
2gG^{ik}=\partial _m \partial _n (\tilde{g}^{ik}
\tilde{g}^{mn}-\tilde{g}^{im} \tilde{g}^{kn})+ L^{ik}~,
\end{equation}
where the term $L^{ik}$  contain the first derivatives only. The
expression
\begin{equation}
U^{ik}=\partial _ m \partial _ n (\tilde{g}^{ik}
\tilde{g}^{mn}-\tilde{g}^{im} \tilde{g}^{kn})~,
\end{equation}
Fock interpreted as gravitational density energy-momentum
pseudo-tensor of weight equal to $+2$. Due to field equations it equal
to $L^{ik}$ and coincides with well-known Landau-Lifshitz
pseudo-tensor.

In RTG  we may to receive the corresponding tensor density,
replacing ordinary derivations to covariant derivations relative
Minkowski metric $\gamma ^{ik}$
\begin{equation}
T_g^{ik}=\frac{1}{\sqrt{-\gamma}}D_m D_n (\tilde{g}^{ik}
\tilde{g}^{mn}-\tilde{g}^{im} \tilde{g}^{kn})~,
\end{equation}
This tensor may be received by means of variational procedure from
the next Lagrangian

\begin{equation}
L=\tilde R(g^{ik})+\frac{1}{\sqrt{-f}} R_{ijkl}(f_{mn}) \tilde
g^{ik}\tilde g^{jl}~.
\end{equation}
where $\tilde {g}^{ik}=\tilde {f}^{ik}+k\tilde {\psi}^{ik}$,
$f_{ik}$ is background metric with nonzero curvature tensor, which
is believed equal to zero after variation. Analogously approach
was used in \cite{Chern}to receive the energy-momentum tensor of
conformal-invariant scalar field.

 Using Fock energy-momentum tensor, let us to calculate the energy and
momentum density for finding solution. For energy density  we have

\begin{equation}
T_g^{00}= -\frac{1}{r^2}[(\frac{R^4}{r^2})_r + 2\frac{R^2}{r}
({\tau_r}^2 g_{\tau\tau} + {R_r}^2 g_{RR})]_r ~,
\end{equation}
This expression is the sum of static and wave parts. For later we
have

\begin{equation}
T^{00}_w = -\frac{2}{r^2}[(\frac{R^2}{r} {g_{\tau\tau})_r}
{T_u}^2 {f_r}^2 + 2\frac{R^2}{r} g_{\tau\tau}(T_u T_{uu}{f_r}^3 +
{T_u}^2 f_r f_{rr})]~.
\end{equation}
The expression (17) may be present as

\begin{equation}
T^{00}_w = \frac{2{T_u}^2}{r^2}[(3-\frac{m^2}{r^2})
{f_r}^2 - 2Cf_r r (1-\frac{m^2}{r^2})(\frac{(r+m)^2}{(r-m)^2}+{f_r}^2 ({T_u}^2 + 2T_u))]~.
\end{equation}

The analysis of last expression shows that it is a
positive-definite, if $C$ and $T_u$ are positive, but $f_r$ is
negative. As it was finding above this conditions follow from
solutions of equations (7),(8).

Let us to find the expression for momentum density.In Cartesian
coordinates $x^i$ we have
\begin{equation}
T_w^{0i}= \frac{4Cx^i}{r^4}(r^2-m^2){f_r}^2 {T_u}^2
(1+T_u)^2 ~,
\end{equation}
This expression is positive-definite for $C>0$ that testify about
energy transport and, consequently, about a wave character  of
finding solution.

For large values of $r$ we have

\begin{equation}
f_r = -1-\frac{1}{Cr}~,
\end{equation}

\begin{equation}
T_w^{00} = \frac{4C}{r} {T_u}^2 (1+{T_u})^2~,
\end{equation}

\begin{equation}
{T_w^{0i}} = \frac{4Cx^i}{r^2} {T_u}^2 (1+{T_u})^2~.
\end{equation}
The expressions (22),(23) show that energy and momentum are not
transporting at infinity as it had to be, because wave solution
was finding with help the coordinate transformation from
asymptotically-flat metric.

Note that the asymptotical expression for retarded argument coincides
with corresponding expression, using by Fock [4] in wave zone
\begin{equation}
u = t-(r+2m \ln{\frac{r}{r_0}})~,
\end{equation}
whence follow that $C=1/2m$.

So, we may to conclude that Hilbert energy-momentum tensor play the
role of gravitational current in the field equations although Fock
tensor describes the energy characteristics of gravitational field.

\section{ CONCLUSION}

The existence of spherically-symmetrical wave solutions and as a whole  nonstatic spherically-symmetrical
solutions is in a contradiction with ordinary physical interpretation of Birkhoff
theorem in GR. In RTG such solutions have a physical sense as far as the temporal coordinate of the
Minkowski space-time has it. Essentially that this wave solution possesses positive-defined
energy and momentum densities.

Although the receiving  solution has  enough formal character, it illustrates the possibility of
existence of spherical gravitational waves. A total solution must include in the interior one
and the matching of these solutions.

\end{document}